\documentclass[12pt]{article}
\usepackage{epsfig}
\usepackage{amsmath}
\usepackage{amssymb}
\usepackage{graphicx}
\usepackage{lscape}
\usepackage {subfigure}
\usepackage{hyperref, cite}
\def\be{\begin{equation}}
\def\ee{\end{equation}}
\def\ba{\begin{array}}
\def\ea{\end{array}}
\def\beqn{\begin{eqnarray}}
\def\eeqn{\end{eqnarray}}

\def\bt{\begin{tabular}}
\def\et{\end{tabular}}
\def\bc{\begin{center}}
\def\ec{\end{center}}
\usepackage{epstopdf}
\begin{document}
\title{ Implications of CP invariants for flavoured hybrid neutrino mass matrix}

\author{Madan Singh$^{*} $\\
\it Department of Physics, M. N. S Government College Bhiwani, Bhiwani,\\
\it Haryana, 127021, India.\\
\it $^{*}$singhmadan179@gmail.com
}
\maketitle
\begin{abstract}
In the present paper, I have re-examined the weak basis invariants at low energies, proposed by C. Jarlskog and Branco et. al, respectively, in their earlier analyses, after confronting them with the assumptions of two  zeros and an equality between arbitrary non-zero elements in the Majorana neutrino mass matrix in the flavoured basis. This particular conjecture is found to be experimentally feasible as shown by S. Dev and D. Raj in their recent work. The present analysis attempts to find the  necessary and sufficient  condition for CP invariance for each  experimentally viable ans\"atz, pertaining to the model along with some important implications.   
\end{abstract}

\section{Introduction}
\section{Introduction}
In the Standard model (SM), CP violation is considered as a potent  tool to explore the flavor sector  \cite{1, 2, 3, 4, 5}, and to probe the signals of New Physics (NP). After the resounding success obtained for K meson decay system, CP violation is  well established in the B meson system as well, owing to the efforts made at  $e^{+} e^{-}$ B factories with their detectors BaBar (SLAC) and Belle(KEK). At present, Cabibbo-Kobayashi-Maskawa (CKM) \cite{6} mechanism appears to be  a only  way
 to understand CP violation in the  SM. This CKM picture of CP violation is well supported by the precision measurements of sin2$\beta$ from the CP asymmetry in the decay B$\rightarrow \psi$ K, as well as by the reconstruction of   unitarity triangle through global fits by various well known groups like Particle Data Group (PDG) \cite{7}, CKMfitter \cite{8}, UTfit \cite{9}  and HFAG \cite{10}. The CP violation  in the quark sector stems from the irremovable  phase $\delta$, which appears in  Cabibbo-Kobayashi-Maskawa (CKM) mixing matrix \cite{6}.  On the contrary, SM does not allow the CP violation in the lepton sector. However, in the spirit of quark-lepton symmetry, it is natural to expect an entirely analogous mechanism to arise in the lepton sector, leading to leptonic CP violation. 
      The  observation of neutrino oscillation,  \cite{11, 12, 13, 14}, in this regard, provided the first evidence that CP may be violated in the lepton sector apart from the fact that neutrinos do mix as well as are massive.   This  further triggers  a sudden spurt of activity on the theoretical as well as experimental front to identify the possibilities beyond the standard model, where leptonic CP violation would be observed.
      The fact that all three neutrino mixing angles ($\theta_{12}, \theta_{23}, \theta_{13}$) have been, measured to a reasonably good degree of precision \cite{15},  only empower our prospects of finding CP violation  in the future experiments. 
      
       In contrast to the  quark case, the parameterization of CP violation in the lepton sector is complicated by several factors. In particular, if the neutrinos are Majorana particles then in comparison to the case of CKM matrix, the corresponding leptonic Pontecorvo-Maki-Nakagawa-Sakata (PMNS) mixing matrix has two additional phases apart from an analogous CKM phase $\delta$.  In the standard parameterisation,  leptonic mixing matrix $\rm{V}$ is given by \cite{16, 17}      
\begin{small}
\begin{equation*}
\rm{V= \left(
\begin{array}{ccc}
 c_{12}c_{13}& s_{12}c_{13}& s_{13}e^{-i\delta} \\
-c_{12}s_{23}s_{13}e^{i\delta}-s_{12}c_{23} & -s_{12}s_{23}s_{13}e^{i\delta}+c_{12}c_{23} & s_{23}c_{13}\\
 -c_{12}c_{23}s_{13}e^{i\delta}+s_{12}s_{23}& -s_{12}c_{23}s_{13}e^{i\delta}-c_{12}s_{23}& c_{23}c_{13} \\
\end{array}
\right)}
\end{equation*} 
\begin{equation*}              
\times \rm{\left(
\begin{array}{ccc}
 1 & 0& 0 \\
0 & e^{i\rho} & 0\\
0& 0& e^{i\sigma} \\
\end{array}
 \right).}
\end{equation*}
\end{small}
Here, $c_{ij} = \cos \theta_{ij}$, $s_{ij}= \sin \theta_{ij}$ for $i, j=1, 2, 3$. $\delta$  denotes a Dirac type phase,  while $\rho, \sigma$ denote  Majorana type CP  phases.
The Dirac type CP  phase  is
expected to be measured in the experiments with superbeams and neutrino beams
from neutrino factories or indirectly through the area of the unitarity triangles defined
for the leptonic sector, while the  Majorana type CP phases will contribute to lepton number violating (LNV) processes like neutrinoless double beta decay.  Therefore, it is expected that neutrino physics could be useful for investigating the leptonic CP violation at low energies apart from having profound
implications for the physics of the early universe.\\
In the flavoured basis, where charged lepton mass matrix is diagonal, Majorana neutrino matrix contains all the information regarding the leptonic CP violation at low energy. To facilitate any specific predictions in this regard, it is convenient to restrict the number of free parameters of  mass matrix. As an example  some elements of neutrino mass matrix can be considered either zero or equal \cite{18,19,20,21} or  some co-factors of neutrino mass matrix to be either zero or equal \cite {19,22,23,24}. Among these possibilities,  simultaneous existence of texture zeros  and an equal elements (or cofactors), also known as hybrid texture, is one of the most notable and rigorously studied in the literature. Not only they reduce the number of free parameters to an appreciable extent compare with texture zero, but also can be naturally realized through flavour symmetry. Recently,  S. Dev and D. Raj \cite{25},  have suggested a new possibility  pertaining to the hybrid  texture comprising two zero elements, and an equality between the non-zero elements.  In contrast to the hybid model with one texture zero and an equality, suggested in Ref.\cite{20,21}, the number of experimentally viable possibilities for new hybrid model, are found to be comparatively lesser (only eight out of total forty two), and  thus  more predictive. 

Nonetheless, like any other texture zero model,  the hybrid textures  are not invariant under weak basis transformation, implying that a given set of  zeros and an equality,  which appears in a certain weak basis (WB) may not be present or may appear in different entries in another WB, while leading to the same physics.
 In view of this, it is always instructive  to examine any specific flavor
model in a basis independent manner. For that purpose, CP invariants  are largely considered as an important tool 
 to investigate the CP properties both in the quark and the leptonic sector. In the literature some special attention has been paid on invariants, owing to their suitability for the analysis of specific ans\"atze of neutrino mass matrix  without even need to diagonalise it. In this regard, S. Dev et al, have studied the WB invariants for the  models with two zeros  \cite{26}, and one zero with a vanishing minor \cite{27} in neutrino mass matrix, respectively, aiming to find the CP properties in lepton sector.\\
 In the present analysis, I shall re-examine the weak basis invariants for  the new possibilty of
 hybrid texture  comprising two zeros and an equality between the non-zero elements in the flavoured basis, and find the CP conditions in lepton sector.
 The interest in the choosen model stems from the fact that a  choice of two zeros and an additional equality in the neutrino mass matrix  
reduces the number of free parameters  to six. Hence it 
is more predictive  as compared to the models with two texture zeros, one zero with a vanishing minor and other hybrid
textures.  
 \\
   
 In order to derive the leptonic CP condition, it is essential to find the WB invariants in terms of the Majorana neutrino mass matrix. To this end,  I shall first define the  WB invariants at low energy \cite{28, 29, 30}, which must  be zero for CP invariance in leptonic sector. The non-zero value signals towards the CP violation. The relevance of CP odd WB invariants in the analysis of the hybrid textures is due to the fact that assumption of two zeros and an equality between the arbitrary elements lead to a decrease in the number of the independent CP violating phases. The number of CP odd invariants coincides with the number of CP-violating phases which arise in the lepton mixing of charged  weak current, after all lepton masses have been diagonalised.  B. Yu and S. Zhou \cite{31} carried out a numerical analysis to show the minimal set of CP-odd invariants, which lead to the CP conservation, are three, and hence the number is not accidental. \\
  The invariant, which  is sensitive to Dirac phase only, is given as
\begin{equation}\label{eq1}
\rm{I1=Tr[m_{\nu } m_{\nu }^{\dagger}, m_{l}m_{l}^{\dagger}]^{3},}
\end{equation}
where, $\rm{m_{\nu}}$ and $\rm{m_{l}}$ are Majorana neutrino mass matrix and charged lepton  mass matrix, respectively. The invariant $\rm{I1}$ is  analogous to  the $\rm{Tr[m_{d} m_{d }^{\dagger}, m_{u}m_{u}^{\dagger}]^{3}}$  \cite{28} in the quark sector with three generations.  The computation of CP violation through $\rm{I1}$ is possible only in the "flavoured basis". Therefore, from  Eq.(\ref{eq1}), 
\begin{equation*}
\rm{I1=-6i\Delta_{e\mu}\Delta_{\mu\tau}\Delta_{\tau e}Im(h_{12}h_{23}h_{31})},
\end{equation*}
 where,  
 $\rm{\Delta_{ij}=m_{j}^{2}-m_{i}^{2}}$, ij run over the pairs, $e\mu,\mu\tau, \tau e$. The argument of  the product, $ \rm{h_{12}h_{23}h_{31}}$, is  responsible for defining the CP properties  for Dirac neutrinos.  The quantity $ \rm{h (\equiv m_{\nu}m_{\nu}^{\dagger})}$, is a hermitian matrix, where 
\begin{eqnarray*}
\rm{h_{11}=|m_{ee}|^{2}+|m_{e\mu}|^{2}+|m_{e\tau}|^{2}},\\
\rm{h_{12}=m_{ee} m_{e\mu}^{*}+m_{e\mu} m_{\mu\mu}^{*}+m_{e\tau} m_{\mu \tau}^{*}},\\
\rm{h_{13}=m_{ee} m_{e\tau}^{*}+m_{e\mu} m_{\mu\tau}^{*}+m_{e\tau} m_{\tau \tau}^{*},}\\
\rm{h_{21}=m_{e\mu} m_{ee}^{*}+m_{\mu\mu} m_{e\mu}^{*}+m_{\mu\tau} m_{e \tau}^{*},}\\
\rm{h_{22}=|m_{e\mu}|^{2}+|m_{\mu \mu}|^{2}+|m_{\mu\tau}|^{2},}\\
\rm{h_{23}=m_{e\mu} m_{e\tau}^{*}+m_{\mu\mu} m_{\mu\tau}^{*}+m_{\mu\tau} m_{\tau \tau}^{*},}\\
\rm{h_{31}=m_{e\tau} m_{ee}^{*}+m_{\mu\tau} m_{e\mu}^{*}+m_{\tau\tau} m_{e \tau}^{*},}\\
\rm{h_{32}=m_{e\tau} m_{e\mu}^{*}+m_{\mu\tau} m_{\mu\mu}^{*}+m_{\tau\tau} m_{\mu \tau}^{*},}\\
\rm{h_{33}=|m_{e\tau}|^{2}+|m_{\mu \tau}|^{2}+|m_{\tau\tau}|^{2}},\\
\end{eqnarray*}
 define the nine matrix elements. The  invariant is sensitive to the Dirac type CP phase $\delta$, and hence depends on lepton number conserving (LNC) process like neutrino oscillations. The vanishing of imaginary part of I1 implies $sin\delta$ = 0, and thus leads to Dirac type CP invariance. The relationship between  $\rm{I1}$ and  Jarlskog rephasing invariant, $\rm{J_{CP}}$ for lepton sector is established  in \cite{28}.  
 The representation of $\rm{J_{CP}}$ in terms of $\rm{sin \delta}$ is found  as
\begin{equation}\label{eq2}
\rm{J_{CP}= \frac{1}{8}s_{2(12)}s_{2(23)}s_{2(13)}c_{13} \sin\delta},
\end{equation}
where, $\rm{s_{2(ij)}\equiv sin2\theta_{ij}}$, $\rm{i,j=1, 2, 3}$.\\
 Also $\rm{J_{CP}}$ is an 'invariant function' of mass matrices, and is related to the mass matrices as
\begin{eqnarray}\label{eq3}
&&
\rm{det C=-2J_{CP} \Delta_{e\mu}\Delta_{\mu\tau}\Delta_{\tau e}  \Delta_{12} \Delta_{23}\Delta_{31}},
\end{eqnarray} 
where, $\rm{C \equiv i[m_{\nu}m_{\nu}^{\dagger}, m_{\l}m_{l}^{\dagger}]}$, and,
$\Delta_{12}, etc$ are analogue to the $\Delta_{ij}$ as defined earlier. 
The commutator C, is by definition, hermitian and traceless. Thus eigen values are real. In fact they are measurable, even though C itself is not a measurable. The determination of any traceless $3\times 3$ may be computed from trace of the third power of the matrix, i.e., $ \rm{detC=\frac{1}{3}Tr(C^{3})}$, therefore I have
$ \rm{det C=-\frac{i}{3}Tr[m_{\nu}m_{\nu}^{\dagger}, m_{\l}m_{l}^{\dagger}]^{3}}$,
which is valid for any traceless 3 $\times$ 3 matrix.
Therefore from Eqs.(\ref{eq3}), I get
\begin{equation}\label{eq4}
\rm{I_{1}= -6iJ_{CP} \Delta_{e\mu}\Delta_{\mu\tau}\Delta_{\tau e}  \Delta_{12} \Delta_{23}\Delta_{31}}.
\end{equation}
The above relation shows  that $\rm{I_{1}}$ is directly proportional to $sin\delta$.\\
 Using Eqs. (\ref{eq2}, \ref{eq4}), I arrive at
\begin{eqnarray}\label{eq5}
\rm{J_{CP}=\frac{Im(h_{12}h_{23}h_{31})}{\Delta_{sol} \Delta_{atm}^{2}}},
\end{eqnarray}
where, $\rm{\Delta_{sol}\equiv \Delta_{12}}$ and $\rm{\Delta_{atm} \equiv \Delta_{23}\simeq \Delta_{31}} $, are solar and atmospheric neutrino mass squared differences.
\\
The other two invariants, $\rm{I2}$ and $\rm{I3}$, are sensitive to both Dirac as well as Majorana type CP phases. The  invariant $\rm{I2}$ is given as
\begin{equation}\label{eq6}
\rm{I2=Im( Tr[m_{l } m_{l}^{\dagger} m_{\nu}^{*}m_{\nu}m_{\nu}^{*}m_{l}^{T}m_{l}^{*}m_{\nu} ])}.
\end{equation}
The above invariant was computed, for the first time, to derive the necessary and sufficient condition for CP invariance in the framework of two  neutrinos \cite{29}. In this framework, CP violation  can occur only due to Majorana type phase. In the flavoured basis, one can re-write  $\rm{I2}$ as a function  of the elements of $m_{\nu}$, 
\begin{eqnarray}\label{eq7}
&&
\rm{I2= m_{e}^{4}m_{ee}H_{11}+m_{\mu}^{4}m_{\mu\mu}H_{22}+m_{\tau}^{4}m_{\tau \tau}H_{33}}
 \nonumber  \\
 && \quad\quad  \rm{+m_{e}^{2}m_{\mu}^{2}m_{e\mu}(H_{12}+H_{21}) +m_{\mu}^{2}m_{\tau}^{2}m_{\mu\tau}(H_{23}+H_{32})}\nonumber  \\
&& \quad\quad \rm{+m_{e}^{2}m_{\tau}^{2}m_{e\tau}(H_{13}+H_{31})},
\end{eqnarray} 
where, H is a complex matrix, and its elements are given as
\begin{equation*}
\rm{H_{11}=m_{ee}^{*}h_{11}+m_{e\mu}^{*}h_{21}+m_{e\tau}^{*}h_{31}},
\end{equation*}
\begin{equation*}
\rm{H_{12}=m_{ee}^{*}h_{12}+m_{e\mu}^{*}h_{22}+m_{e\tau}^{*}h_{32}},
\end{equation*}
\begin{equation*}
\rm{H_{13}=m_{ee}^{*}h_{13}+m_{e\mu}^{*}h_{23}+m_{e\tau}^{*}h_{33}},
\end{equation*}
\begin{equation*}
\rm{H_{21}=m_{e\mu}^{*}h_{11}+m_{\mu\mu}^{*}h_{21}+m_{\mu\tau}^{*}h_{31}},
\end{equation*}
\begin{equation*}
\rm{H_{22}=m_{e\mu}^{*}h_{12}+m_{\mu\mu}^{*}h_{22}+m_{\mu\tau}^{*}h_{32}},
\end{equation*}
\begin{equation*}
\rm{H_{23}=m_{e\mu}^{*}h_{13}+m_{\mu\mu}^{*}h_{23}+m_{\mu\tau}^{*}h_{33}},
\end{equation*}
\begin{equation*}
\rm{H_{31}=m_{e\tau}^{*}h_{11}+m_{\mu\tau}^{*}h_{21}+m_{\tau\tau}^{*}h_{31}},
\end{equation*}
\begin{equation*}
\rm{H_{32}=m_{e\tau}^{*}h_{12}+m_{\mu\tau}^{*}h_{22}+m_{\tau\tau}^{*}h_{32}},
\end{equation*}
\begin{equation*}
\rm{H_{33}=m_{e\tau}^{*}h_{13}+m_{\mu\tau}^{*}h_{23}+m_{\tau\tau}^{*}h_{33}}.
\end{equation*}
From the above, it is apparent that all the elements are complex. Therefore H depends on both Dirac as well as Majorana type CP  phases.\\
On substituting  $\rm{m_{\nu}m_{\nu}^{\dagger}}$ with $\rm{m_{\nu}(m_{l}m_{l}^{\dagger})^{*}m_{\nu}^{\dagger}}$ in Eq.(\ref{eq1}),  $\rm{I_{3}}$ can  be trivially expressed as, 
\begin{equation}\label{eq8}
\rm{I3\equiv Tr[m_{\nu}m_{l}^{T}m_{l}^{*}m_{\nu}^{\dagger}, m_{l}m_{l}^{\dagger}]^{3}}.
\end{equation}
The computation  of  above invariant  predicts CP violation for  three or more generation of Majorana neutrinos even in the limit of complete neutrino mass degeneracy \cite{29,30}. This is  contrary to  the case of Dirac neutrinos, where in the limit of exact degeneracy it is well known that there is no CP violation or physical lepton mixing, for an arbitrary number of generations.\\ 
  In the choosen basis, invariant $\rm{I3}$ can be written as
\begin{equation*}
\rm{I3=-6i\Delta_{e\mu}\Delta_{\mu\tau}\Delta_{\tau e}Im(h_{12}^{'}h_{23}^{'}h_{31}^{'})},
\end{equation*}
where, $\rm{arg(h_{12}^{'}h_{23}^{'}h_{31}^{'}}$) is complex phase responsible for CP violation.\\
 Similar to h, $\rm{h^{'}}$ is also hermitian matrix, 
 and its elements are given as 
\begin{equation*}
\rm{h_{11}^{'}=m_{e}^{2}|m_{ee}|^{2}+m_{\mu}^{2}|m_{e\mu}|^{2}+m_{\tau}^{2}|m_{e\tau}|^{2},}\\
\end{equation*}
\begin{equation*}
\rm{h_{12}^{'}=m_{e}^{2}m_{ee} m_{e\mu}^{*}+m_{\mu}^{2}m_{e\mu} m_{\mu\mu}^{*}+m_{\tau}^{2}m_{e\tau} m_{\mu \tau}^{*},}\\
\end{equation*}
\begin{equation*}
\rm{h_{13}^{'}=m_{e}^{2}m_{ee} m_{e\tau}^{*}+m_{\mu}^{2}m_{e\mu} m_{\mu\tau}^{*}+m_{\tau}^{2}m_{e\tau} m_{\tau \tau}^{*},}\\
\end{equation*}
\begin{equation*}
\rm{h_{21}^{'}=m_{e}^{2}m_{e\mu} m_{ee}^{*}+m_{\mu}^{2}m_{\mu\mu} m_{e\mu}^{*}+m_{\tau}^{2}m_{\mu\tau} m_{e \tau}^{*},}\\
\end{equation*}
\begin{equation*}
\rm{h_{22}^{'}=m_{e}^{2}|m_{e\mu}|^{2}+m_{\mu}^{2}|m_{\mu \mu}|^{2}+m_{\tau}^{2}|m_{\mu\tau}|^{2},}\\
\end{equation*}
\begin{equation*}
\rm{h_{23}^{'}=m_{e}^{2}m_{e\mu} m_{e\tau}^{*}+m_{\mu}^{2}m_{\mu\mu} m_{\mu\tau}^{*}+m_{\tau}^{2}m_{\mu\tau} m_{\tau \tau}^{*},}\\
\end{equation*}
\begin{equation*}
\rm{h_{31}^{'}=m_{e}^{2}m_{e\tau} m_{ee}^{*}+m_{\mu}^{2}m_{\mu\tau} m_{e\mu}^{*}+m_{\tau}^{2}m_{\tau\tau} m_{e \tau}^{*},}\\
\end{equation*}
\begin{equation*}
\rm{h_{32}^{'}=m_{e}^{2}m_{e\tau} m_{e\mu}^{*}+m_{\mu}^{2}m_{\mu\tau} m_{\mu\mu}^{*}+m_{\tau}^{2}m_{\tau\tau} m_{\mu \tau}^{*},}\\
\end{equation*}
\begin{equation*}
\rm{h_{33}^{'}=m_{e}^{2}|m_{e\tau}|^{2}+m_{\mu}^{2}|m_{\mu \tau}|^{2}+m_{\tau}^{2}|m_{\tau\tau}|^{2},}\\
\end{equation*}
where, $\rm{m_{e}, m_{\mu}}$ and $\rm{m_{\tau}}$ denote the  electron, muon and tau neutrino, respectively.\\
\section{CP invariants for eight experimentally viable hybrid textures}
 Among the  fourty-two phenomenolically possible ansazte of hybrid texture with two zeros and an equality between the non-zero elements, only eight are found to be viable in the light of current experimental data at 3$\sigma$ confidence level (CL).    The analysis in \cite{25} shows that  all the viable ansatze    give a specific prediction on neutrino mass ordering. The ansazte belonging to class A1 and A2, respectively, allow only normal mass ordering , while ansazte belonging to class C allow only inverted mass ordering. Following the classification scheme of \cite{25},  all the viable ans\"atze have been encapsulated in Table \ref{tab1}.

Among the  viable ans\"atze, there are only five independent ones.
 The ansatze corresponding to each pair, $(\rm{A1^{IV}, A2^{IV}}), (\rm{A1^{V}, A2^{V}})$ and $(\rm{A1^{VI}, A2^{VI}})$,  exhibit the similar phenomenological implications, and are  related via permutation symmetry.  The symmetry  corresponds to the  simultaneous exchange  of  2-3 rows and 2-3 columns of $\rm{m_{\nu}}$. The corresponding permutation matrix can be given by
\begin{equation}\label{eq9}
 \rm{P_{23} = \left(
\begin{array}{ccc}
    1& 0& 0 \\
  0 & 0 & 1\\
  0& 1& 0 \\
\end{array}
\right)}.
\end{equation}
Using  permutation symmetry, it is trivial to find the  following relations among the neutrino oscillation parameters,
\begin{equation}\label{eq10}
\rm{\theta_{12}^{X}=\theta_{12}^{Y}, \ \
\theta_{23}^{X}=90^{\circ}-\theta_{23}^{Y},\ \
\theta_{13}^{X}=\theta_{13}^{Y}, \ \ \delta^{X}=\delta^{Y} -180^{\circ}},
\end{equation}
where X and Y denote the pair of  aforementioned ans\"atze  related through a 2-3 permutation symmetry.  On the other hand, ans\"atze $\rm{C^{II}}$ and $\rm{C^{IV}}$ 
 tranform onto themselves independently.\\
\begin{table}[htp]
\begin{center}
\resizebox{8.5cm}{!}{
\begin{tabular}{|c|c|c|c|}
\hline & \textbf{IV} & \textbf{V}&\textbf{VI}\\   
 \hline Class A1 &$\left(
\begin{array}{ccc}
0 & 0 & e \\
 & b & b\\
&  & c \\
\end{array}
\right)$ &$\left(
\begin{array}{ccc}
0 & 0 & e \\
 & b & f\\
&  & b \\
\end{array}
\right)$ &$\left(
\begin{array}{ccc}
0 & 0 & e \\
 & b & c\\
&  & c \\
\end{array}
\right)$ \\

\hline 
\hline & \textbf{IV} & \textbf{V}&\textbf{VI}\\   
 \hline Class A2 &$\left(
\begin{array}{ccc}
0 & d & 0 \\
 & b & b\\
&  & c \\
\end{array}
\right)$ &$\left(
\begin{array}{ccc}
0 & d & 0 \\
 & b & f\\
&  & b \\
\end{array}
\right)$ &$\left(
\begin{array}{ccc}
0 & d & 0 \\
 & b & c\\
&  & c \\
\end{array}
\right)$ \\
\hline 
\hline
&\textbf{II} & \textbf{IV}&  \\
\hline Class C &$\left(
\begin{array}{ccc}
    a & d & a \\
 & 0 & f\\
&  & 0 \\
\end{array}
\right)$&$\left(
\begin{array}{ccc}
    a & d & e \\
 & 0 & e\\
&  & 0 \\
\end{array}
\right)$&\\
\hline

\end{tabular}}
\caption{\label{tab1}The eight viable ans\"{a}tze  of Majorana neutrino mass matrices having two texture zeros and one equality. }

\end{center}
\end{table}

In this section, I compute the WB invariants  in terms of mass matrix elements for all the eight experimentally viable hybrid textures, and consequently find the CP conditions corresponding to each ans\"atz.\\ 
\textbf{Ansatz $\rm{(A1)^{IV}}$}: Using Eq.(\ref{eq1}), it is trivial to obtain,\\
\begin{equation}\label{eq11}
\rm{I1=-6i\Delta_{e\mu}\Delta_{\mu\tau}\Delta_{\tau e}|m_{e\tau}|^{2}|m_{\mu\mu}|^{2}ImQ_{\rm{(A1)^{IV}}}},
\end{equation}
where, $\rm{Q_{\rm{(A1)^{IV}}}= m_{\tau\tau}m_{\mu\mu}^{*}}$.\\
Using the above equation, $\rm{J_{CP}}$, which is a measurable neutrino oscillation parameter, can be given as
\begin{equation*}
\rm{J_{CP}=\frac{|m_{e\tau}|^{2}|m_{\mu\mu}|^{2}ImQ_{\rm{(A1)^{IV}}}}{\Delta_{sol} \Delta_{atm}^{2}}}.
\end{equation*}
From the expression, it is convenient to directly compute  $\rm{J_{CP}}$ from the elements of $\rm{m_{\nu}}$  without even need to diagonalize the $\rm{m_{\nu}}$. Interestingly, condition  $\rm{m_{\mu \mu}=m_{\tau\tau}}$, leads to $\rm{J_{CP}=0}$ or equivalently sin$\delta$=0. Hence $\rm{m_{\mu \mu}=m_{\tau\tau}}$ is essential to find the  CP invariance relevant for Lepton number conserving process (LNC). 
\\
Using Eqs. (\ref{eq7}) and (\ref{eq8}), the expressions for invariants $\rm{I2}$ and $\rm{I3}$, sensitive to both Dirac and Majorana phases,  are found as
\begin{equation}\label{eq12}
\rm{I2=\Delta_{\mu\tau}^{2}|m_{\mu\mu}|^{2} ImQ_{\rm{(A1)^{IV}}}},
\end{equation}
 and,
\begin{equation}\label{eq13}
\rm{I3=-6i\Delta_{e\mu}\Delta_{\mu\tau}\Delta_{\tau e} m_{\mu}^{2}m_{\tau}^{4}  |m_{e\tau}|^{2}|m_{\mu\mu}|^{2}ImQ_{\rm{(A1)^{IV}}}}.
\end{equation}
respectively.
From the above discussion, it is clear that CP violation is possible if the phases associated with the $\rm{m_{\mu\mu}}$ and $\rm{m_{\tau\tau}}$ are not finely tuned.

More apparently, CP invariance condition is allowed for \textbf{ansatz $\rm{(A1)^{IV}}$}  if 
and only if 
\begin{equation}\label{eq14}
\rm{arg(m_{\mu\mu})=arg(m_{\tau \tau})}.
\end{equation}\\

\textbf{Ansatz $\rm{(A1)^{V}}$}:  Similarly, using Eq.(\ref{eq1}), one can find invariant I1 as 

\begin{equation}\label{eq15}
\rm{I1=-6i\Delta_{e\mu}\Delta_{\mu\tau}\Delta_{\tau e}|m_{e\tau}|^{2}ImQ_{\rm{(A1)^{V}}}},
\end{equation}
where, $\rm{Q_{\rm{(A1)^{V}}}= m_{\mu\mu}^{2}(m_{\mu\tau}^{*})^{2}}$.\\
The Jarlskog rephasing invariant parameter, can be found  in terms of mass matrix elements as
\begin{equation}\label{eq16}
\rm{J_{CP}=\frac{|m_{e\tau}|^{2}|m_{\mu\mu}|^{2}ImQ_{\rm{(A1)^{IV}}}}{\Delta_{sol} \Delta_{atm}^{2}}}.
\end{equation}
Therefore, condition $\rm{m_{\mu\mu}=m_{\mu\tau}}$, implies $\rm{J_{CP}=0}$.\\
The other two invariants I2 and I3 can be derived using Eqs.((\ref{eq7}), (\ref{eq8}))
\begin{equation}\label{eq17}
\rm{I2=\Delta_{\mu\tau}^{2}|m_{e\tau}|^{2} ImQ_{\rm{(A1)^{V}}}},
\end{equation}
 and,
\begin{equation}\label{eq18}
\rm{I3=-6i\Delta_{e\mu}\Delta_{\mu\tau}\Delta_{\tau e} m_{\mu}^{2}m_{\tau}^{4}  |m_{e\tau}|^{2}ImQ_{\rm{(A1)^{V}}}}.
\end{equation}
The CP invariance condition can be found as
\begin{equation}\label{eq19}
\rm{arg(m_{\mu\mu})=arg(m_{\mu \tau})}.
\end{equation}
The  mismatch between the elements $\rm{m_{\mu \mu}}$ and $\rm{m_{\mu\tau}}$  lead to CP violation in lepton sector.\\
\textbf{Ansatz} $\rm{(A1)^{VI}}$: Using Eqs.(\ref{eq1}), it is found that\\
\begin{equation}\label{eq20}
\rm{I1=-6i\Delta_{e\mu}\Delta_{\mu\tau}\Delta_{\tau e}|m_{e\tau}|^{2}|m_{\tau\tau}|^{2}ImQ_{\rm{(A1)^{VI}}}},
\end{equation}
 where, $\rm{Q_{\rm{(A1)^{VI}}}= m_{\mu \mu} m_{\tau\tau}^{*}}$.\\
 Using Eq.(\ref{eq5}), $\rm{J_{CP}}$ can be written as
 \begin{equation}\label{eq21}
 \rm{J_{CP}=\frac{|m_{e\tau}|^{2}|m_{\tau\tau}|^{2}ImQ_{\rm{(A1)^{IV}}}}{\Delta_{sol} \Delta_{atm}^{2}}}.
 \end{equation}
 As found in ans\"atz $\rm{A1^{IV}}$, Eq. (\ref{eq21}) provides $\rm{J_{CP}=0}$ for  $\rm{m_{\mu \mu} = m_{\tau\tau}}$.
 In addition phase relation for CP invariance is similar to ans\"atz $\rm{A1^{IV}}$.\\ 
 The other invariants I2 and I3  can be derived using Eqs.(\ref{eq7}) and (\ref{eq8}), 
\begin{equation}\label{eq22}
\rm{I2=\Delta_{\mu\tau}^{2}|m_{\tau\tau}|^{2} ImQ_{\rm{(A1)^{VI}}}},
\end{equation}
 and,
\begin{equation}\label{eq23}
\rm{I3=-6i\Delta_{e\mu}\Delta_{\mu\tau}\Delta_{\tau e} m_{\mu}^{2}m_{\tau}^{4}  |m_{e\tau}|^{2}|m_{\tau\tau}|^{2}ImQ_{\rm{(A1)^{VI}}}}.
\end{equation}
Similarly, one can find the expressions for WB invariants for ans\"atze $\rm{A2^{IV}, A2^{V}}$ and $\rm{A2^{VI}}$ from  $\rm{A1^{IV}, A1^{V}}$ and $\rm{A1^{VI}}$, respectively, by using the $\mu-\tau$ exchange permutation symmetry, as mentioned earlier.\\
The condition for CP invariance is similar for $\rm{A2^{IV}}$ ($\rm{A2^{VI}}$) as found for $\rm{A1^{IV}}$ ($\rm{A1^{VI}}$).\\

\textbf{Ansatz} $\rm{C^{II}}$: The invariant I1 can be derived by using Eq.(\ref{eq1})\\
\begin{equation}\label{eq24}
\rm{I1=-6i\Delta_{e\mu}\Delta_{\mu\tau}\Delta_{\tau e}(|m_{e\mu}|^{2}-|m_{e\tau}|^{2})|m_{ee}|^{2}ImQ_{C^{II}}},
\end{equation}
 where, $\rm{Q_{C^{II}}\equiv m_{\mu\tau} m_{e\mu}^{*}}$.\\
 The Jarlskog rephasing invariant parameter $\rm{J_{CP}}$ can be written as
 \begin{equation}\label{eq25}
 \rm{J_{CP}= \frac{(|m_{e\mu}|^{2}-|m_{e\tau}|^{2})|m_{ee}|^{2}ImQ_{C^{II}}}{\Delta_{sol} \Delta_{atm}^{2}}}.
 \end{equation}
 From the above equation, it is explicit that $\rm{J_{CP}}=0$ might be due to $|\rm{m_{e\mu}}|=|\rm{m_{e\tau}}|$. Keeping in mind the texture zero condition for ansatz $\rm{C^{II}}$, it can be inferred that  Dirac type CP symmetry is related to $\mu-\tau$ reflection symmetry (i.e. $\rm{m_{e\mu}=\pm m_{e\tau}, m_{\mu\mu}=m_{\tau\tau}}$ ) \cite{32,33}. The other possibility for CP invariance arises if $\rm{m_{\mu\tau} =m_{e\mu}}$ is considered.\\
Interestingly, it is found from the above equation that $\rm{J_{CP}}$ and $\rm{|m_{ee}|}$ are strongly related to each other, i.e.  $\rm{J_{CP} \propto |m_{ee}|^{2}}$. This implies the parabolic relation between these experimentally measurable parameters, and further point  out that the absence of  neutrinoless double beta decay  is  related to $sin\delta=0$ if other possibilities of CP invariance is relaxed.\\
 The remaining invariants I2 and I3 can be written as
\begin{equation}\label{eq26}
\rm{I2=2\Delta_{e\mu}\Delta_{\tau e}|m_{ee}|^{2}ImQ_{C^{II}}},
\end{equation}
and,
\begin{equation}\label{eq27}
\rm{I3=-6i\Delta_{e\mu}\Delta_{\mu\tau}\Delta_{\tau e} m_{e}^{4}(m_{\mu}^{2}|m_{e\mu}|^{2}-m_{\tau}^{2}|m_{e\tau}|^{2})|m_{ee}|^{2}ImQ_{C^{II}}}.
\end{equation}
As already mentioned that CP invariants, I2 and I3,  are sensitive to  CP properties associated to Majorana nature of neutrinos. In the scenario of $\mu-\tau$ reflection symmetry, invariants I2 and I3 are non-zero and hence CP symmetry can be broken. This is contrary to the case of invariant I1, where CP symmetry can be preserved. Therefore, it can be stated that, for ans\"atz $\rm{C^{II}}$, CP violation is attributed to only Majorana type phases ($\rho, \sigma$) (and independent of Dirac type phase $\delta$) , if $\mu-\tau$ reflection symmetry is assumed.    \\

The CP invariance condition can be given as
\begin{equation}\label{eq28}
\rm{arg(m_{\mu\tau})=arg(m_{e \mu})}.
\end{equation}

\textbf{Ansatz} $\rm{C^{IV}}$: Using Eq.(\ref{eq1}), invariant I1 can be written as \\ 
\begin{equation}\label{eq29}
\rm{I1=-6i\Delta_{e\mu}\Delta_{\mu\tau}\Delta_{\tau e}(|m_{e\mu}|^{2}-|m_{e\tau}|^{2})|m_{e\tau}|^{2}ImQ_{C^{IV}}},
\end{equation}
 where,  $\rm{Q_{C^{IV}}\equiv m_{ee} m_{e\mu}^{*}}$.\\
 The Jarlskog rephasing invariant  can be written as
 \begin{equation}\label{eq30}
 \rm{J_{CP}=\frac{(|m_{e\mu}|^{2}-|m_{e\tau}|^{2})|m_{e\tau}|^{2}ImQ_{C^{IV}}}{\Delta_{sol} \Delta_{atm}^{2}}}.
 \end{equation}
 Again for $\rm{m_{ee}= m_{e\mu}}$ , $\rm{J_{CP}=0}$. Similar to ans\"atz $\rm{C^{II}}$, $\rm{J_{CP}}=0$ can also hold if $\mu-\tau$ reflection symmetry is assumed.
 \\
 The other invariants I2 and I3 can be written as
\begin{equation}\label{eq31}
\rm{I2=2\Delta_{e\mu}\Delta_{\tau e}|m_{e\tau}|^{2}ImQ_{C^{IV}}},
\end{equation}
and,
\begin{equation}\label{eq32}
\rm{I3=-6i\Delta_{e\mu}\Delta_{\mu\tau}\Delta_{\tau e} m_{e}^{4}(m_{\mu}^{2}|m_{e\mu}|^{2}-m_{\tau}^{2}|m_{e\tau}|^{2})|m_{e\tau}|^{2}ImQ_{C^{IV}}}.
\end{equation}
 Similar  arguments holds for the ansatz $\rm{C^{IV}}$ if $\mu-\tau$ reflection symmetry is considered, as  discussed for ansatz $\rm{C^{II}}$. 

The CP invariance condition is given as
\begin{equation}\label{eq33}
\rm{arg(m_{ee})=arg(m_{e\mu})}.
\end{equation}
From the above discussion, it is inferred that condition of either additional equality or texture zero automatically lead to CP invariance. The maximum number of allowed choices of  texture zeros and an equal elements, needed to  find the CP violation, are restricted to that considered in present case of hybrid textures. 
In addition, from  Eqs.(\ref{eq14}), (\ref{eq19}), (\ref{eq28}) and (\ref{eq33}),  it is sufficient to assume one of the elements of mass matrix  to be real, for the hybrid texture model to be  CP invariant. For illustration, if an element  $\rm{m_{ee}}$ is  considered real, or zero, which essentially implies that neutrinos are Dirac particle. The assumption  simultaneously lead to vanishing imaginary part of  $\rm{ m_{e\mu}}, \rm{m_{\mu \tau}}, \rm{m_{\mu\mu}}$ and $\rm{m_{\tau\tau}}$, respectively [ see Eqs. (\ref{eq14}), (\ref{eq19}), (\ref{eq28}),(\ref{eq33})].
 Hence the non-observation of neutrinoless beta decay in future experiments could imply that current hydrid texture is   CP invariant. \\
The Dirac CP phase, which is sensitive to the CP violation in  neutrino oscillation is contained in WB invariant I1, while invariants I2 and I3 are measures of Majorana type CP phases, contribute  to CP violation in neutrinoless double beta decay. However, it is explicit from the analysis,  that all the three CP violating phases are not independent and there is only one physical phase in all the phenomenologically viable ansatze with two texture zeros and an equality between non-zero elements. However Dirac and Majorana CP type phases can not be distinguished in a concerned hybrid texture of Majorana mass matrix.

\section{Summary  and conclusion}
To summarize the analysis, I have computed the three weak basis invariants in terms of neutrino mass matrix elements  at low energy for   hybrid textures of Majorana neutrinos with two texture zeros and an equality between the non-zero mass matrix elements in the flavoured basis. 
In the analysis, some useful equalities between the different neutrino mass matrix elements are found to have some profound connection with CP invariance pertaining to both lepton number conserving (LNC) and  lepton number violating (LNV) processes.   Similar to the earlier analyses by S. Dev et al \cite{26,27}, it is maintained that there is only one physical phase which describes the CP properties in present model, while Dirac and Majorana phases can not be extracted without considering certain assumptions.

\section*{Conflicts of Interest}
The author declares that there are no conflicts of interest regarding the publication of this paper.

\section*{Acknowledgment}

The author would like to thank the Principal, M. N. S Government college, Bhiwani , for providing necessary facilities to work.  Thanks to  S. Dev and D. Raj for bringing forward  the  research work pertaining to new possibilty of hybrid texture.  \\

\newpage

\end{document}